\documentclass[]{interact}

\usepackage{epstopdf}
\usepackage[caption=false]{subfig}
\usepackage{amsmath}

\usepackage[numbers,sort&compress]{natbib}
\bibpunct[, ]{[}{]}{,}{n}{,}{,}
\renewcommand\bibfont{\fontsize{10}{12}\selectfont}
\makeatletter
\def\NAT@def@citea{\def\@citea{\NAT@separator}}
\makeatother

\theoremstyle{plain}

\theoremstyle{definition}

\theoremstyle{remark}

\begin{document}

\articletype{ARTICLE TEMPLATE}

\title{Phase discontinuities induced scintillation enhancement: coherent vortex beams propagating through weak oceanic turbulence}

\author{
\name{Hantao Wang\textsuperscript{a}, Huajun Zhang\textsuperscript{a}, Mingyuan Ren\textsuperscript{a},  Jinren Yao\textsuperscript{a} and Yu Zhang\textsuperscript{a}\thanks{CONTACT Yu Zhang. Email: zhangyuhitphy@163.com}}
\affil{\textsuperscript{a}School of Physics, Harbin Institute of Technology, No.92, Xidazhi Road, Harbin 150001, China}
}

\maketitle

\begin{abstract}
Under the impact of an infinitely extended edge phase dislocation, optical vortices (screw phase dislocations) induce scintillation enhancement. The scintillation index of a beam consisting of two Gaussian vortex beams with ${\pm{1}}$ topological charges through weak oceanic turbulence is researched via derivation and phase screen simulation. Different combinations of two types of phase discontinuities can be obtained by changing the overlapping degree and the phase difference of two coherent Gaussian vortex beams. The scintillation indexes for them verify that the formation condition of the phenomenon is the coexistence of two types of phase discontinuities. And the enhanced scintillation index can be several orders of magnitude larger than that of a plane wave under weak perturbation (Rytov variance). This phenomenon could be useful for both optical vortex detection and perturbation measurement.
\end{abstract}


\section{Introduction}
When a beam propagates through turbulence, the fluctuation of refractive index will modify the complex amplitude of the beam and result in optical turbulence. If the turbulence is strong enough, some points with zero-amplitude and undetermined phase will appear in pairs which called branch points. Tracing around a point counterclockwise, there will be a continuous $2\pi$ phase increase (decrease) which makes the point be a phase singularity with positive (negative) charge. Besides, each pair of branch points with opposite topological charges are the origins of a phase dislocation line which is called branch cuts. Nye and Berry initially observed this phenomenon in wave trains and introduced the conception into optics \cite{nye1974dislocations}. Since then, phase discontinuities have been studied widely in theory and experiment, especially in the areas of atmosphere turbulence and adaptive optics. Fried and Vaughn elaborated the difference between the phase discontinuity at the dark rings of the Airy diffraction pattern and branch cuts that the phase discontinuity step size of the former is $\upi$ , or  ${\upi}+2k\pi$ rather than $2\pi$ of the latter \cite{fried1992branch}. Based on these obvious contrasts, phase discontinuities can be divided into three main types:  infinitely extended edge dislocation, screw dislocations, and limited edge dislocation \cite{basistiy1995optical}. The first one is a classical dislocation in far-field diffraction patterns and the laser beams with transverse cavity modes \cite{lugiato1990instabilities}. The third one can be regarded as a mixed type and it is unstable under the influence of perturbations. By contrast, for screw dislocations, more novel characters drive researchers to pay more attention to its evolution behavior in free space or under different perturbations.

Initially the propagation dynamic of screw dislocations, also called optical vortices in optics \cite{coullet1989optical}, affected by several optical structures (e.g., optical vortices with the same or opposite topological, background field with phase gradient or amplitude gradient) in free space has been investigated \cite{ahluwalia2005evolution}. Then with the dramatic development of optical communication system and Lidar in atmosphere and ocean, the propagation behavior of optical vortices in turbulence becomes a popular topic \cite{cheng2016propagation}. The focus is switched to the dynamical behavior in external perturbations. For optical communication, the ultimate objective is reducing the influence of turbulence and promoting the stability of optical signal \cite{ren2016orbital}. Several types of laser beams with optical vortices have been shown to be less affected by turbulence when compared with non-vortex beams \cite{2016Vortex,li2017influence,liu2013experimental}. However, for Lidar to measure turbulence, optical vortex becomes a new parameter to characterize some aspects of turbulence. Ref. \cite{voitsekhovich1998density} derived the theoretical expression of dislocation density in various turbulent and propagation conditions. Then, numerical experiments were carried out to prove that singularity density increases slowly with the increase of Rytov index which represents the intensity of turbulence \cite{rao2008statistics}. Therefore, the singularity density may be used to depict the intensity of turbulence \cite{oesch2009aggregate}. 

Looking back on these researches, those who focused on the evolution behaviors of screw dislocations affected by initial optical fields or external perturbations accelerate the development of various applications of optical vortices in adaptive optics \cite{2009The}, optical vortex field manipulation \cite{he1995direct} and optical measurement \cite{vadnjal2013measurement}. Further researches on the joint influence aimed to reduce the impact of external perturbations through exploring appropriate forms of initial optical fields \cite{aksenov2012increase}. Thus far, the interaction of a pair of branch points through distributed turbulence has been unknown, especially when they are infinitesimally close together \cite{2011Orbital}. And the discussion about the combination impact of screw dislocations and infinitely extended edge dislocations on optical fields under perturbations has not been published yet. Therefore, in this paper, referring to the structure of branch points and the stability of optical vortices \cite{freund1999critical}, a pair of coherent Gaussian vortex beams with opposite single-charged optical vortices were considered. An infinitely extended edge dislocation can be obtained by fully overlapping these two beams. When these two beams are partial overlapping, both an infinitely extended edge dislocation and screw dislocations coexist except for the condition that the phase difference of two beams is $\upi$. Besides, scintillation, as a phenomenon describing the intensity fluctuations of a beam propagating through turbulence, was chosen to reflect the behavior of two types of phase discontinuities. That is not only because the evolution of phase discontinuities can induce the intensity variation of a beam, but also because scintillation index has been widely applicated in characterizing the intensity of turbulence.

In Section 2, we derived the scintillation index of the beam through oceanic turbulence and, to verify the applicability of analytical derivation, discussed that in two cases: completely overlapping (only an infinitely extended edge dislocation exists) and completely separating (only screw dislocations exist). Next, in Section 3, we considered the partial overlapping condition containing both two types of phase discontinuities to present the formation conditions and the characters of enhanced scintillation. Then in Section 4, we used phase screens method to demonstrate the phenomenon mentioned in Section 3 and verified the theoretical analysis. In the end, we discussed the results of Section 3 and Section 4 in Section 5.

\section{The scintillation of coherent beams with screw dislocations or an infinitely extended edge dislocation}

For the simplicity of discussion, the scintillation of the optical field consisting of two coherent Gaussian vortex beams with ${\pm{1}}$ topological charges in any overlap conditions is presented. The expression can be written as
\begin{equation}
U\left( {\boldsymbol{\rho '},z = 0} \right) = {U_ + }\left( {\boldsymbol{\rho '} - \boldsymbol{d},z = 0} \right) + {U_ - }\left( {\boldsymbol{\rho '} + \boldsymbol{d},z = 0} \right)\exp \left( {i\phi } \right),
\end{equation}
where
\begin{equation}
{U_ \pm }\left( {\boldsymbol{\rho '},z = 0} \right) = {E_ \pm }\exp \left( { - \frac{{{{\boldsymbol{\rho '}}^2}}}{{{\sigma ^2}}}} \right)\left( {{\rho '_x} \pm i{\rho '_y}} \right),
\end{equation}
and $\boldsymbol{\rho '} \equiv \left( {{\rho '_x},{\rho '_y}} \right)$ is a two dimensional vector at the plane $z=0$, $\phi$ is the initial phase difference of two beams and ${E_ \pm }$ is the electric field amplitude of beams with ${\pm{1}}$ topological charges. To simplify the calculation, the amplitude of two beams are assumed to be the same value $E_0$. The degree of overlap is determined jointly by the distance of the centers of two beams $2d$ and the waist width of the Gaussian background beam envelope $\sigma$. When $d$ is equal to zero or approaches to infinite, the state can be regarded as fully overlap or complete separation, respectively. Except these two limiting cases, an infinitely extended edge dislocation and two screw dislocations with opposite sign coexist in different distributions. For simplification and symmetry, the direction of separation is chosen to be x-axis and the origin of coordinates is established at the middle point of the centers of two beams.

The general expression of the scintillation index of a beam through random media is described as follows \cite{andrews2005laser}:
\begin{equation} \label{scintillation}
\sigma _I^2 = \frac{{\left\langle {{I^2}\left( {\boldsymbol{\rho },z} \right)} \right\rangle }}{{{{\bigl\langle {I\left( {\boldsymbol{\rho },z} \right)} \bigr\rangle }^2}}} - 1,
\end{equation}
where ${I\left( {\boldsymbol{\rho },z} \right)}$ and ${{I^2}\left( {\boldsymbol{\rho },z} \right)}$ denote the instantaneous intensity and the square of the instantaneous intensity at the plane $z$ with transverse coordinates $\boldsymbol{\rho} \equiv \left( {{\rho_x},{\rho_y}} \right)$. $\left\langle  \cdot  \right\rangle $ represents the ensemble average. It is obvious that the second moment and fourth moment of optical field should be derived, respectively, before calculating the scintillation index. Therefore, we first present the second-order statistics of optical field.

In order to simultaneously characterize the variance of intensity and the evolution behavior of phase discontinuities, the extended Huygens-Fresnel principle and the cross-spectral density method are used. The cross-spectral density function at the receiving plane can be expressed as \cite{huang2014evolution}
\begin{align} \label{cross}
 W\left( {{\boldsymbol{\rho }_1},{\boldsymbol{\rho }}_2,z} \right) =
& \frac{{{k^2}E_0^2}}{{4{{\rm{\upi }}^2}{z^2}}}\mathrm{\iint}{U^ * }\left( {{\boldsymbol\rho '_1},0} \right)U\left( {{\boldsymbol\rho '_2},0} \right)\Bigl\langle {\exp \left[ {{\psi ^ * }\left( {{{\boldsymbol\rho'_1}},{\boldsymbol\rho_1},z} \right)} \right.\left. { + \psi \left( {{{\boldsymbol\rho '_2}},{\boldsymbol\rho_2},z} \right)} \right]} \Bigr\rangle \nonumber\\
&\times \exp \left\{ { - \frac{{ik}}{{2z}}\left[ {{{\left( {{{\boldsymbol\rho '_1}} - {\boldsymbol\rho _1}} \right)}^2} + {{\left( {{{\boldsymbol\rho '_2}} - {\boldsymbol\rho_2 }} \right)}^2}} \right]} \right\}{\rm{d}}{{\boldsymbol\rho '_1}}{\rm{d}}{{\boldsymbol\rho '_2}},
\end{align}
where $k = {{2{\rm{\upi }}} \mathord{\left/{\vphantom {{2{\rm{\upi }}} \lambda }} \right.\kern-\nulldelimiterspace} \lambda }$ is the wave number that is related to the wavelength $\lambda$, $\psi$ is the random part of the complex phase of a spherical wave induced by turbulence, the asterisk $*$ denotes the complex conjugate. The ensemble average part in Eq. (\ref{cross}) contains the Bessel function of the first kind and zero order. And Ref. \cite{Shirai2003Mode} discussed the required condition of its simplification. The approximation can be expressed as  
\begin{align} \label{approximation1}
&\Bigl\langle {\exp \left[ {{\psi ^ * }\left( {{{\boldsymbol\rho '_1}},{\boldsymbol\rho _1},z} \right) + \psi \left( {{{\boldsymbol\rho '_2}},{\boldsymbol\rho _2},z} \right)} \right]} \Bigr\rangle \nonumber\\
 \approx &\exp \biggl\{ { - {k^2}zT\left[ {{{\left( {{{\boldsymbol\rho '_1}} - {{\boldsymbol\rho '_2}}} \right)}^2} + \left( {{{\boldsymbol\rho '_1}} - {{\boldsymbol\rho '_2}}} \right)\left( {{\boldsymbol\rho_1} - {\boldsymbol\rho_2}} \right) + {{\left( {{\boldsymbol\rho _1} - {\boldsymbol\rho_2}} \right)}^2}} \right]} \biggr\},
\end{align}
and
\begin{align}
T = \frac{{{{\rm{\upi }}^2}}}{3}\int_0^\infty  {{\kappa ^3}} {\Phi _{\rm{n}}}\left( \kappa  \right){\rm{d}}\kappa .
\end{align}
Here, $\Phi _{\rm{n}}\left( \kappa  \right)$ is the wide-range Prandtl/Schmidt number power spectrum of refractive index fluctuations. It can be expressed as the linear combination of temperature spectrum $\Phi _{\rm{T}}\left( \kappa  \right)$, the salinity spectrum $\Phi _{\rm{S}}\left( \kappa  \right)$ and the co-spectrum $\Phi _{\rm{TS}}\left( \kappa  \right)$ in the form of \cite{yao2019wide}
\begin{align}
{\Phi _{\rm{n}}}\left( \kappa  \right) = {A^2}{\Phi _{\rm{T}}}\left( \kappa  \right) + {B^2}{\Phi _{\rm{S}}}\left( \kappa  \right) + 2AB{\Phi _{{\rm{TS}}}}\left( \kappa  \right),
\end{align}
$A$ and $B$ are the linear coefficients related to average temperature ${\left\langle {\rm{T}} \right\rangle }$ and average salinity concentration ${\left\langle {\rm{S}} \right\rangle }$. Each spectrum can be written as \cite{yao2019wide}
\begin{align}\label{spectrumi}
{\Phi _i}\left( \kappa  \right) =& \left[ {1 + 21.61{{\left( {\kappa \eta } \right)}^{0.61}}{c_i}^{0.02} - 18.18{{\left( {\kappa \eta } \right)}^{0.55}}{c_i}^{0.04}} \right]\nonumber\\
&\times \frac{1}{{4\pi }}\beta {\varepsilon ^{ - \frac{1}{3}}}{\kappa ^{ - \frac{{11}}{3}}}{\chi _i}\exp \Bigl[ { - 174.90{{\left( {\kappa \eta } \right)}^2}{c_i}^{0.96}} \Bigr].{\kern 10pt}i \in \left\{ {{\rm{T}},{\rm{S}},{\rm{TS}}} \right\}
\end{align}
By the way, owing to the distribution of $\Phi _{\rm{n}}\left( \kappa  \right)$, although the approximation does not satisfy the requirements in Ref. \cite{Shirai2003Mode}, Eq. (\ref{approximation1}) is still valid with the restrictive conditions that the intensity of turbulence is extremely weak and the transverse scale of the beam is small. This required condition should also be satisfied in the subsequent parts. In Eq. (\ref{spectrumi}) the Kolmogorov microscale $\eta$ is defined as \cite{yao2020spatial}
\begin{align}
\eta  = {v^{{3 \mathord{\left/
 {\vphantom {3 4}} \right.
 \kern-\nulldelimiterspace} 4}}}{\varepsilon ^{{{ - 1} \mathord{\left/
 {\vphantom {{ - 1} 4}} \right.
 \kern-\nulldelimiterspace} 4}}},
\end{align}
where $v$ is the momentum diffusivity and $\varepsilon$ is the energy dissipation rate. The dimensionless parameters ${c_i}\left( i\in \left\{ {{\rm{T}},{\rm{S}},{\rm{TS}}} \right\} \right)$ are \cite{yao2020spatial}
\begin{align}
{c_{\rm{T}}} = {0.072^{{4 \mathord{\left/
 {\vphantom {4 3}} \right.
 \kern-\nulldelimiterspace} 3}}}\beta_0 P{r^{ - 1}},{\kern 2pt} {c_{\rm{S}}} = {0.072^{{4 \mathord{\left/
 {\vphantom {4 3}} \right.
 \kern-\nulldelimiterspace} 3}}}\beta_0 S{c^{ - 1}},
{\kern 2pt} {{c_{{\rm{TS}}}} = {{0.072}^{{4 \mathord{\left/
 {\vphantom {4 3}} \right.
 \kern-\nulldelimiterspace} 3}}}\beta_0 \frac{{Pr + Sc}}{{2PrSc}}},
\end{align}
where $Pr$ is the Prandtl number, $Sc$ is the Schmidt number and $\beta_0$ is the Obukhov-Corrsin constant that is equal to 0.72. In Eq. (\ref{spectrumi}) ${\chi_i}\left( i\in \left\{ {{\rm{T}},{\rm{S}},{\rm{TS}}} \right\} \right)$ are the ensemble-averaged variance dissipation that can be defined by \cite{yao2020spatial}
\begin{align}
&{\chi _{\rm{T}}} = {K_{\rm{T}}}{\left( {\frac{{d\left\langle {\rm{T}} \right\rangle }}{{dz}}} \right)^2},{\kern 2pt}
{\chi _{\rm{S}}} = {K_{\rm{S}}}{\left( {\frac{{d\left\langle {\rm{S}} \right\rangle }}{{dz}}} \right)^2} = \frac{{{d_r}}}{{{H^2}}}{\chi _{\rm{T}}},\nonumber\\
&{{\chi _{{\rm{TS}}}} = \frac{{{K_{\rm{T}}} + {K_{\rm{S}}}}}{2}\left( {\frac{{d\left\langle {\rm{T}} \right\rangle }}{{dz}}} \right)\left( {\frac{{d\left\langle {\rm{S}} \right\rangle }}{{dz}}} \right) = \frac{{1 + {d_r}}}{{2H}}{\chi _{\rm{T}}}},
\end{align}
where $K_{\rm{T}}$ and $K_{\rm{S}}$ are the eddy diffusivity of temperature and salinity, respectively. The eddy diffusivity ratio $d_r$ is derived from density ratio ${R_\rho } = {{\alpha \left| H \right|} \mathord{\left/
 {\vphantom {{\alpha \left| H \right|} \beta }} \right.
 \kern-\nulldelimiterspace} \beta }$. And it can be written as \cite{yao2020spatial}
\begin{align}
{d_r} \approx \left\{ {\begin{array}{*{20}{c}}
\begin{array}{l}
{R_\rho } + R_\rho ^{0.5}{\left( {{R_\rho } - 1} \right)^{0.5}},\\
1.85{R_\rho } - 0.85,\\
0.15{R_\rho },
\end{array}&\begin{array}{l}
{R_\rho } \ge 1,\\
0.5 \le {R_\rho } < 1,\\
{R_\rho } < 0.5.
\end{array}
\end{array}} \right.
\end{align}
The $H$ represents the temperature-salinity gradient ratio, $\alpha$ and $\beta$ are the thermal expansion coefficient and saline contraction coefficient, respectively. The intensity of turbulence can be determined only by $\varepsilon$, $H$, $\chi_{\rm{T}}$, ${\left\langle {\rm{T}} \right\rangle}$ and ${\left\langle {\rm{S}} \right\rangle }$. For intuitive characterization, we prefer using Rytov variance $\sigma _R^2$ (the scintillation for a plane wave) as a recognized indicator to describe the fluctuation conditions of oceanic turbulence. Therefore, the analytical expression of Rytov variance is derived as follows:
\begin{align}\label{rytovvariance}
\sigma _R^2 &= 8{\upi ^2}{k^2}L\int_0^1 {\int_0^\infty  {\kappa {\Phi _n}\left( \kappa  \right)} } \left[ {1 - \cos \left( {\frac{{L{\kappa ^2}\xi }}{k}} \right)} \right]{\rm{d}}\kappa {\rm{d}}\xi \nonumber\\
 &= {A^2}\sigma _{R{\rm{T}}}^2 + {B^2}\sigma _{R{\rm{S}}}^2 + 2AB\sigma _{R{\rm{TS}}}^2,
\end{align}
with
\begin{align}
{\sigma_{Ri}^2} = & \upi {k^2}L{\beta _0}{\varepsilon ^{ - \frac{1}{3}}}{\chi _i}{\eta ^{\frac{5}{3}}}\sum\limits_{j = 1}^3 {{b_j}} {\left( {174.90{c_i}^{0.96}} \right)^{\frac{5}{6} - {a_j}}}\left\{ {\frac{{36}}{{36a_j^2 - 24{a_j} - 5}}} \right.\nonumber\\
&\times \Gamma \left( {\frac{7}{6} + {a_j}} \right) + \frac{6}{{6{a_j} - 11}}{\left( {1 + \theta _i^{ - 2}} \right)^{\frac{5}{{12}} - \frac{{{a_j}}}{2}}}\Gamma \left( { - \frac{5}{6} + {a_j}} \right)\nonumber\\
&\times \Biggl\{ {\cos \left[ {\left( {\frac{5}{6} - {a_j}} \right){\mathop{\rm arccot}\nolimits} \left( {{\theta _i}} \right)} \right] + {\theta _i}\sin \left[ {\left( {\frac{5}{6} - {a_j}} \right){\mathop{\rm arccot}\nolimits} \left( {{\theta _i}} \right)} \right]} \Biggr\}\Biggr\},\nonumber\\
&i\in \left\{ {{\rm{T}},{\rm{S}},{\rm{TS}}} \right\},
\end{align}
and
\begin{align}
{b_j} = \left( {\begin{array}{*{20}{c}}
1&{21.61{c_i}^{0.02}}&{ - 18.18{c_i}^{0.04}}
\end{array}} \right),{\kern 10pt} {a_j} = \left( {\begin{array}{*{20}{c}}
0&{\frac{{0.61}}{2}}&{\frac{{0.55}}{2}}
\end{array}} \right),
\end{align}
where $\theta _i = \left( {174.90{c_i}^{0.96}} \right){{{\eta ^2}k} \mathord{\left/
 {\vphantom {{{\eta ^2}k} L}} \right.
 \kern-\nulldelimiterspace} L}$ and $\Gamma$ is the Gamma function.

Return to the derivation of cross-spectral density, the analytical expression of $T$ based on ${\Phi _i}\left( \kappa  \right)$ is obtained. Similar to the derivation of Rytov variance in Eq. (\ref{rytovvariance}), $T$ is also able to transform into the linear combination of ${T_i}\left( {i \in \left\{ {{\rm{T}},{\rm{S}},{\rm{TS}}} \right\}} \right)$ that can be written as
\begin{align}
{T_i} = \frac{\upi }{{24}}{\beta _0}{\varepsilon ^{ - \frac{1}{3}}}{\chi _i}{\eta ^{ - \frac{1}{3}}}\sum\limits_{j = 1}^3 {{b_j}} {\left( {174.90{c_i}^{0.96}} \right)^{ - \frac{1}{6} - {a_j}}}\Gamma \left( {\frac{1}{6} + {a_j}} \right).
\end{align}
Note that, for simplicity, the parameters associated with ${\Phi _{\rm{n}}}\left( \kappa  \right)$ in the rest of this paper are just presented by the components associated with ${\Phi _i}\left( \kappa  \right)$. Then, according to the integral formula \cite{gradshteyn2014table}
\begin{align}
 \int_{ - \infty }^\infty  {{x^n}\exp \left( { - p{x^2} + 2qx} \right)} {\rm{d}}x = n!\exp \left( {\frac{{{q^2}}}{p}} \right)\sqrt {\frac{{\rm{\upi }}}{p}} {\left( {\frac{q}{p}} \right)^n}\sum\limits_{k = 0}^{\left\lfloor {{n \mathord{\left/
 {\vphantom {n 2}} \right.
 \kern-\nulldelimiterspace} 2}} \right\rfloor } {\frac{1}{{\left( {n - 2k} \right)!\left( k \right)!}}{{\left( {\frac{p}{{4{q^2}}}} \right)}^k}},
\end{align}
the result of Eq. (\ref{cross}) is shown as follows
\begin{align}
 W\left( {{\boldsymbol\rho_1},{\boldsymbol\rho_2},z} \right) =& \frac{{{k^2}E_0^2}}{{4{z^2}}}\frac{1}{{p_1^2p_2^2}}\exp \left( { - \frac{{2{d^2}}}{{{\sigma ^2}}}} \right)\exp \Bigl[ { - {k^2}zT{{\left( {{\boldsymbol\rho_1} - {\boldsymbol\rho _2}} \right)}^2}} \Bigr]\exp \left( { - ik\frac{{\boldsymbol\rho^2_1 - \boldsymbol\rho^2_2}}{{2z}}} \right)\nonumber\\
&\times \exp \left( {\frac{{q_{1y}^2}}{{{p_1}}} + \frac{{q_{2y}^2}}{{{p_2}}}} \right)\left( {{S_{ +  + }} + {S_{ -  - }} + {C_{ +  - }} + {C_{ -  + }}} \right),
\end{align}
with
\begin{align}
 {S_{ \pm  \pm }}\left( {{\boldsymbol\rho_1},{\boldsymbol\rho_2},z} \right) =& \exp \left[ {\frac{{{{\left( {{q_{1x}} \pm {D_{1x}}} \right)}^2}}}{{{p_1}}} + \frac{{{{\left( {{q_{2x}} \pm {D_{1x}} \pm {D_{2x}}} \right)}^2}}}{{{p_2}}}} \right]\Biggl\{ {\left( {{q_{1x}} \pm {D_{1x}} \pm d{p_1} \mp i{q_{1y}}} \right)}\nonumber\\
& \times \left( {{q_{2x}} \pm {D_{1x}} \pm {D_{2x}} \pm d{p_2} \pm i{q_{2y}}} \right){\rm{ + }}{k^2}zT{\rm{ + }}{k^2}zT\nonumber\\
&\times\left[{\frac{{1}}{{{p_2}}} {{\left( {{q_{2x}} \pm {D_{1x}} \pm {D_{2x}}} \right)}^2} + \frac{{q_{2y}^2}}{{{p_2}}} \pm d\left( {{q_{2x}} \pm {D_{1x}} \pm {D_{2x}} \mp i{q_{2y}}} \right)} \right] \Biggr\},
\end{align}
\begin{align}
 {C_{ \pm  \mp }}\left( {{\boldsymbol{\rho }_1},{\boldsymbol{\rho }_2},z} \right) =& \exp \left( { \pm i\phi } \right)\exp \left[ {\frac{{{{\left( {{q_{1x}} \pm {D_{1x}}} \right)}^2}}}{{{p_1}}} + \frac{{{{\left( {{q_{2x}} \mp {D_{1x}} \pm {D_{2x}}} \right)}^2}}}{{{p_2}}}} \right]\nonumber\\
&  \times \Biggl\{ {\left( {{q_{1x}} \pm {D_{1x}} \pm d{p_1} \mp i{q_{1y}}} \right)}\left( {{q_{2x}} \mp {D_{1x}} \pm {D_{2x}} \mp d{p_2} \mp i{q_{2y}}} \right){\rm{ + }}{k^2}zT\nonumber\\
&\times \left[ {\frac{1}{{{p_2}}}{{\left( {{q_{2x}} \mp {D_{1x}} \pm {D_{2x}} \mp i{q_{2y}}} \right)}^2} \mp d\left( {{q_{2x}} \mp {D_{1x}} \pm {D_{2x}} \mp i{q_{2y}}} \right)} \right] \Biggr\},
\end{align}
where
\begin{align}
 &{p_1} = \frac{1}{{{\sigma ^2}}} + {k^2}zT + \frac{{ik}}{{2z}},{\kern 2pt}
{p_2} = \frac{1}{{{\sigma ^2}}} + {k^2}zT - \frac{{ik}}{{2z}} - \frac{{{k^4}{z^2}{T^2}}}{{{p_1}}},
{\kern 2pt}
{D_{1x}} =  - \frac{d}{{{\sigma ^2}}}, {\kern 2pt} {D_{2x}} =  - \frac{{{k^2}zTd}}{{{\sigma ^2}{p_1}}},\nonumber\\
 &{\boldsymbol{q}_1} = \frac{1}{2}{k^2}zT\left( {{\boldsymbol\rho_1} - {\boldsymbol\rho_2}} \right) + \frac{{ik}}{{2z}}{\boldsymbol{\rho }_2},{\kern 2pt}
{\boldsymbol{q}_2} =  - \frac{1}{2}{k^2}zT\left( {{\boldsymbol\rho_1} - {\boldsymbol\rho_2}} \right) - \frac{{ik}}{{2z}}{\boldsymbol{\rho }_1} + \frac{{{\boldsymbol{q}_1}{k^2}zT}}{{{p_1}}},
\end{align}
${\boldsymbol{q}_1} \equiv \left( {{q_{1x}},{q_{1y}}} \right)$ and ${\boldsymbol{q}_2} \equiv \left( {{q_{2x}},{q_{2y}}} \right)$. ${S_{ \pm  \pm }}$ represents the cross-spectral density of single beams with $\pm{1}$ topological charges and ${C_{ \pm  \mp }}$ is the cross-term of two beams. The $\left\langle I \right\rangle$ in Eq. (\ref{scintillation}) can be obtained when ${\boldsymbol{\rho }_1} = {\boldsymbol{\rho }_2}$. In addition, the evolution behavior of phase discontinuities can be obtained based on spectral degree of coherent which is defined as \cite{mandel1995optical}
\begin{align}
\mu \left( {{\boldsymbol{\rho }_1},{\boldsymbol{\rho }_2},z} \right) = \frac{{W\left( {{\boldsymbol{\rho }_1},{\boldsymbol{\rho }_2},z} \right)}}{{\sqrt {I\left( {{\boldsymbol{\rho }_1},z} \right)I\left( {{\boldsymbol{\rho }_2},z} \right)} }}.
\end{align}
And the position of optical vortex is determined by \cite{gbur2003coherence}
\begin{align}
{\mathop{\rm Re}\nolimits} \left[ {\mu \left( {{\boldsymbol{\rho }_1},{\boldsymbol{\rho }_2},z} \right)} \right] = 0,
{\kern 2pt}
{\mathop{\rm Im}\nolimits} \left[ {\mu \left( {{\boldsymbol{\rho }_1},{\boldsymbol{\rho }_2},z} \right)} \right] = 0,
\end{align}
where $\rm{Re}$ denotes the real part and $\rm{Im}$ represents the imaginary part.

After completing the derivation of the second-order statistics, we set about discussing the fourth-order statistics of optical field. The general fourth-order cross-coherence function at receive plane can be expressed in the form of \cite{andrews2005laser}
\begin{align} \label{forth_order}
 W\left( {{\boldsymbol{\rho }_1},{\boldsymbol{\rho }_2},{\boldsymbol{\rho }_3},{\boldsymbol{\rho }_4},z} \right) = &\frac{{{k^4}}}{{16{\upi ^4}{z^4}}}\mathrm{\iiiint}U\left( {{\boldsymbol\rho '_1},0} \right){U^ * }\left( {{\boldsymbol\rho '_2},0} \right)U\left( {{\boldsymbol\rho '_3},0} \right){U^ * }\left( {{\boldsymbol\rho '_4},0} \right)\nonumber\\
&\times \Bigl\langle {\exp \bigl[ {\psi \left( {{\boldsymbol\rho_1},{{\boldsymbol\rho '_1}},z} \right) + {\psi ^ * }\left( {{\boldsymbol\rho_2},{{\boldsymbol\rho '_2}},z} \right)} }  + \psi \left( {{\boldsymbol\rho_3},{{\boldsymbol\rho '_3}},z} \right)\nonumber\\
&{ { + {\psi ^ * }\left( {{\boldsymbol\rho_4},{{\boldsymbol\rho '_4}},z} \right)} \bigr]} \Bigr\rangle \exp \left\{ {\frac{{ik}}{{2z}}} \right.\left[ {{{\left( {{{\boldsymbol\rho '_1}} - {\boldsymbol\rho_1}} \right)}^2} - {{\left( {{{\boldsymbol\rho '_2}} - {\boldsymbol{\rho }_2}} \right)}^2}} \right.\nonumber\\
&{\left. { + {{\left( {{{\boldsymbol\rho '_3}} - {\boldsymbol{\rho }_3}} \right)}^2} - {{\left( {{{\boldsymbol\rho '_4}} - {\boldsymbol{\rho }_4}} \right)}^2}} \right]} \biggr\}{\rm{d}^2}{{\boldsymbol\rho '_1}}{\rm{d}^2}{{\boldsymbol\rho '_2}}{\rm{d}^2}{{\boldsymbol\rho '_3}}{\rm{d}^2}{{\boldsymbol\rho '_4}},
\end{align}
where
\begin{align}
 &\Bigl\langle {\exp \left[ {\psi \left( {{\boldsymbol{\rho }_1},{\boldsymbol\rho '_1},z} \right) + {\psi ^ * }\left( {{\boldsymbol{\rho }_2},{\boldsymbol\rho '_2},z} \right) + \psi \left( {{\boldsymbol{\rho }_3},{\boldsymbol\rho '_3},z} \right) + {\psi ^ * }\left( {{\boldsymbol{\rho }_4},{\boldsymbol\rho '_4},z} \right)} \right]} \Bigr\rangle \nonumber\\
 = &\exp \Bigl[ {4{E_1}\left( {0,0} \right)}  + {E_2}\left( {{\boldsymbol{\rho }_1} - {\boldsymbol{\rho }_2},{\boldsymbol\rho '_1} - {\boldsymbol\rho '_2}} \right) + {E_2}\left( {{\boldsymbol{\rho }_1} - {\boldsymbol{\rho }_4},{\boldsymbol\rho '_1} - {\boldsymbol\rho '_4}} \right)\nonumber\\
& + {E_2}\left( {{\boldsymbol{\rho }_3} - {\boldsymbol{\rho }_2},{\boldsymbol\rho '_3} - {\boldsymbol\rho '_2}} \right) + {E_2}\left( {{\boldsymbol{\rho }_3} - {\boldsymbol{\rho }_4},{\boldsymbol\rho '_3} - {\boldsymbol\rho '_4}} \right)\nonumber\\
&+ {E_3}\left( {{\boldsymbol{\rho }_1} - {\boldsymbol{\rho }_3},{\boldsymbol\rho '_1} - {\boldsymbol\rho '_3}} \right) + E_3^ *  {\left( {{\boldsymbol{\rho }_2} - {\boldsymbol{\rho }_4},{\boldsymbol\rho '_2} - {\boldsymbol\rho '_4}} \right)} \Bigr],
\end{align}
and
\begin{align} \label{approximation2}
{E_1}(0,0) =  - 2{\upi ^2}{k^2}\int_0^L {\int_0^\infty  {\kappa {\Phi _n}\left( \kappa  \right)} } \rm{d}\kappa \rm{d}z,
\end{align}
\begin{align}\label{approximation3}
{E_2}\left( {{\boldsymbol{\rho }_1},{\boldsymbol{\rho }_2}} \right) = 4{\upi ^2}{k^2}\int_0^L {\int_0^\infty  {\kappa {\Phi _n}\left( {\kappa ,z} \right)} } {J_0}\left\{ {\kappa \left| {\left( {{\boldsymbol{\rho }_1} - {\boldsymbol{\rho }_2}} \right)} \right|} \right\}\rm{d}\kappa \rm{d}z,
\end{align}
\begin{align}\label{approximation4}
{E_3}\left( {{\boldsymbol{\rho }_1},{\boldsymbol{\rho }_2}} \right) = - 4{{\rm{\upi }}^2}{k^2}L\int_0^1 \int_0^\infty  &{{\rm{\kappa }}{\Phi _n}\left( {{\rm{\kappa }},z} \right)}  {J_0}\left\{ {{\rm{\kappa }}\left| {\left( {{\boldsymbol{\rho }_1} - {\boldsymbol{\rho }_2}} \right)} \right|} \right\}\nonumber\\
 &\times \exp \left\{ { - \frac{{iL{{\rm{\kappa }}^2}}}{k}\left( {1 - \xi } \right)\xi } \right\}d\kappa d\xi ,
\end{align}
${J_0}\left( {x} \right)$ is the first kind and zero order of the Bessel function, $\xi  = 1 - {z \mathord{\left/
 {\vphantom {z L}} \right.
\kern-\nulldelimiterspace} L}$ is the normalized distance variable. The analytical derivation of $W\left( {{\boldsymbol{\rho }_1},{\boldsymbol{\rho }_2},{\boldsymbol{\rho }_3},{\boldsymbol{\rho }_4},z} \right)$ is complicated and only $\left\langle {{I^2}} \right\rangle $ is what we really want. Therefore, Eq. (\ref{forth_order}) is reasonable  to be simplified into the form where ${\boldsymbol{\rho }_1} = {\boldsymbol{\rho }_2} = {\boldsymbol{\rho }_3} = {\boldsymbol{\rho }_4} = \boldsymbol{\rho }$. Combining the approximation used in Eq. (\ref{approximation1}) and the simplification method, the calculations of Eqs. (\ref{approximation2}--\ref{approximation4}) are able to transform into
\begin{align}
&{E_1}\left( {0,0} \right) =  - 2{\upi ^2}{k^2}L{T_0},{\kern 2pt}
{E_2}\left( {\boldsymbol{\rho },\boldsymbol{\rho },{\boldsymbol\rho '_1},{\boldsymbol\rho '_2}} \right) \approx 4{\upi ^2}{k^2}L{T_0} - {k^2}LT{\left( {{\boldsymbol\rho '_1} - {\boldsymbol\rho '_2}} \right)^2},\nonumber\\
&{E_3}\left( {\boldsymbol{\rho },\boldsymbol{\rho },{\boldsymbol\rho '_1},{\boldsymbol\rho '_2}} \right) \approx  - 4{{\rm{\upi }}^2}{k^2}L{T_1} + {{\rm{\upi }}^2}{k^2}L{T_2}{\left( {{\boldsymbol\rho '_1} - {\boldsymbol\rho '_2}} \right)^2},
\end{align}
and the analytical expressions of ${T_{0}}$, ${T_{1}}$ and ${T_{2}}$ are presented as follows
\begin{align}
 {T_{0i}} = \frac{1}{{8\pi }}{\beta _0}{\varepsilon ^{ - \frac{1}{3}}}{\chi _i}{\eta ^{\frac{5}{3}}}\sum\limits_{j = 1}^3 {{b_j}} {\left( {174.90{c_i}^{0.96}} \right)^{\frac{5}{6} - {a_j}}}\Gamma \left( { - \frac{5}{6} + {a_j}} \right),
\end{align}
\begin{align}
 {T_{1i}} = &\frac{1}{{4\pi }}{\beta _0}{\varepsilon ^{ - \frac{1}{3}}}{\chi _i}{2^{ - \frac{8}{3}}}\sum\limits_{j = 1}^3 {\left[ {{b_j}{{\left( {2\eta } \right)}^{2{a_j}}}\Gamma \left( {{a_j} - \frac{5}{6}} \right){{\left( {4 \times 174.90{\eta ^2}{c_i}^{0.96} + i\frac{L}{k}} \right)}^{\frac{5}{6} - {a_j}}}{\mkern 1mu} } \right.} \nonumber\\
 &\left. {{\kern 1pt} {\kern 1pt} {\kern 1pt}  \times {\kern 1pt} {}_2{F_1}\left( {\frac{1}{2}, - \frac{5}{6} + {a_j};\frac{3}{2};\frac{{iL}}{{iL + 4k174.90{\eta ^2}{c_i}^{0.96}}}} \right)} \right],
\end{align}
\begin{align}
 {T_{2i}} =& \frac{1}{{4\pi }}{\beta _0}{\varepsilon ^{ - \frac{1}{3}}}{\chi _i}\sum\limits_{j = 1}^3 {{b_j}{\eta ^{2{a_j}}}\Gamma \left( {{a_j} - \frac{5}{6}} \right)} {\kern 1pt} \left[ {\frac{1}{4}{{\left( {4 \times 174.90{\eta ^2}{c_i}^{0.96} + i\frac{L}{k}} \right)}^{ - 1}}} \right.\nonumber\\
 &  \times {\left( {174.90{\eta ^2}{c_i}^{0.96}} \right)^{\frac{5}{6} - {a_j}}} - \frac{{ik}}{{4L}}{\left( {174.90{\eta ^2}{c_i}^{0.96}} \right)^{\frac{5}{6} - {a_j}}} + {4^{{a_j} - \frac{5}{6}}}{L^{ - 1}}\nonumber\\
 & \times {\left( {4 \times 174.90{\eta ^2}{c_i}^{0.96} + i\frac{L}{k}} \right)^{ - \frac{1}{6} - {a_j}}}\left( {ik174.90{\eta ^2}{c_i}^{0.96} + \frac{{6{a_j} - 11}}{{12}}L} \right)\nonumber\\
 & \times {}_2{F_1}\left. {\left( {\frac{1}{2}, - \frac{5}{6} + {a_j};\frac{3}{2};\frac{{iL}}{{iL + 4k174.90{\eta ^2}{c_i}^{0.96}}}} \right)} \right],
\end{align}
where ${}_2{F_1}$ is the hypergeometric function. The analytical expression of $\left\langle {{I^2}} \right\rangle$ is able to obtained in accordance with the procedure used in second-order statistics derivation through onerous calculations. However, the derivation results are too complicated to be presented in this paper. We just substitute these results into the scintillation index and present the numerical results of scintillation index with different situations. 

In order to verify the applicability of the analytical derivation result obtained, we choose two general cases to calculate their scintillation indexes. Here, we discuss the first case corresponding to the condition of $d \to \infty $ that only a screw dislocation exists. In other words, the scintillation of a Gaussian vortex beam with single charged and the evolution behavior of optical vortex are investigated. In this case, there is no additional background filed to influence the propagation of the optical vortex. Therefore, we just expound the scintillation index with different intensities of oceanic turbulence and how an optical vortex evolves. The variables associated to oceanic turbulence are set to be fixed values except $\chi_{\rm{T}}$ to achieve the most convenient way of changing the intensity of oceanic turbulence. The fixed parameters are set as: $\varepsilon  = {10^{ - 2}}{{\rm{m}}^2}{{\rm{s}}^{ - 3}}$, $H =  - 20^\circ {\rm{C}} \cdot {\rm{pp}}{{\rm{t}}^{ - 1}}$, $\left\langle {\rm{T}} \right\rangle  = 15^\circ {\rm{C}}$ and $\left\langle {\rm{S}} \right\rangle  = 34.9{\rm{ppt}}$. On the basis of that, the value ranges of $\chi_{\rm{T}}$ and the parameters of the initial beam should be determined under the consideration of the approximation of ${J_0}\left( {x} \right)$ from power series expansion and the applicability of approximate method used in Rytov variance. So the transverse size of the initial beam is set to be about the order of magnitude of $10^{-3}\rm{m}$, $\chi_{\rm{T}}$ is ranging from $0$ to ${10^{ - 8}}{{\rm{K}}^2}{{\rm{s}}^{ - 1}}$ and the transmission distance is chosen to be $5\rm{m}$. In this case, the Rytov variance is capped at ${2.6\times10^{-5}}$.

\begin{figure}[htb]
\centering\includegraphics[width=0.4\linewidth]{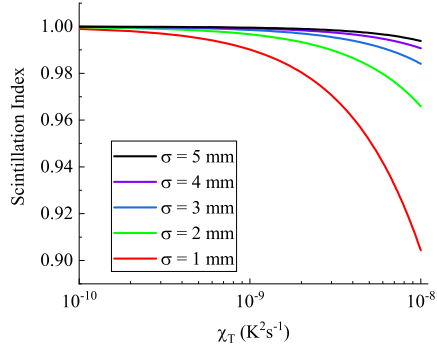}
\caption{The evolution behavior of on-axis scintillation index of Gaussian vortex beam with single charged through oceanic turbulence for different $\sigma$ and $\chi_{\rm{T}}$.}\label{singlebeam}
\end{figure}

The on-axis scintillation indexes of the Gaussian vortex beam whose wavelength is $532\rm{nm}$ with single charged versus the intensity of weak oceanic turbulence are illustrated for different $\sigma$ in Figure \ref{singlebeam}. The variation of scintillation indexes for several values of $\sigma$ are small under perturbation. And the variation range is in agreement with that in Ref. \cite{aksenov2015scintillations}. Therefore, the approximation in Eq. (\ref{approximation1}) is valid in these conditions. The smaller the $\sigma$ is, the steeper the descending portion of the scintillation index curve will be. That is caused by the rapid raise of the on-axis ensemble average intensity that reflects the sensitivity of the beam to oceanic turbulence. The on-axis scintillation index of the Gaussian vortex beam is widely different from that of the Gaussian beam in weak oceanic turbulence for its large scintillation index. But for resemblance, these beams still retain weak response to the variation of the intensity of oceanic turbulence.

\begin{figure}[htb]
\centering\includegraphics[width=0.4\linewidth]{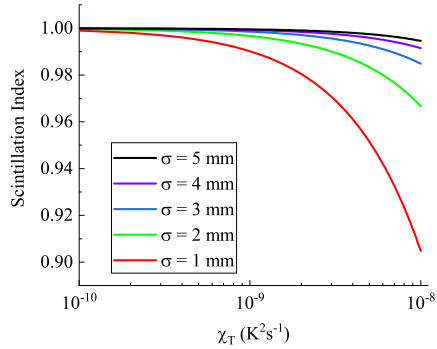}
\caption{The evolution behavior of on-axis scintillation index of a beam that only contains an infinitely extended edge dislocation through oceanic turbulence for different $\sigma$ and $\chi_{\rm{T}}$.}\label{doublebeam}
\end{figure}

The second case is the condition of $d=0$ that only an infinitely extended edge dislocation exists. The optical field of this case can be regarded as one of two mutually orthogonal components of the light field in the first case with the same distribution at $z$ plane. Because of the statistical homogeneity and isotropy of oceanic turbulence, the scintillation index of any component is the same. Therefore, the variation of scintillation index is the same as that in the first case. The evolution behavior of on-axis scintillation index for $d=0$ is presented in Figure \ref{doublebeam}. The trend of the curve is in accord with the prediction before. 

\section{Scintillation enhancement induced jointly by two types of phase discontinuities}

\begin{figure}[htb]
\centering\subfloat{\includegraphics[width=0.4\linewidth]{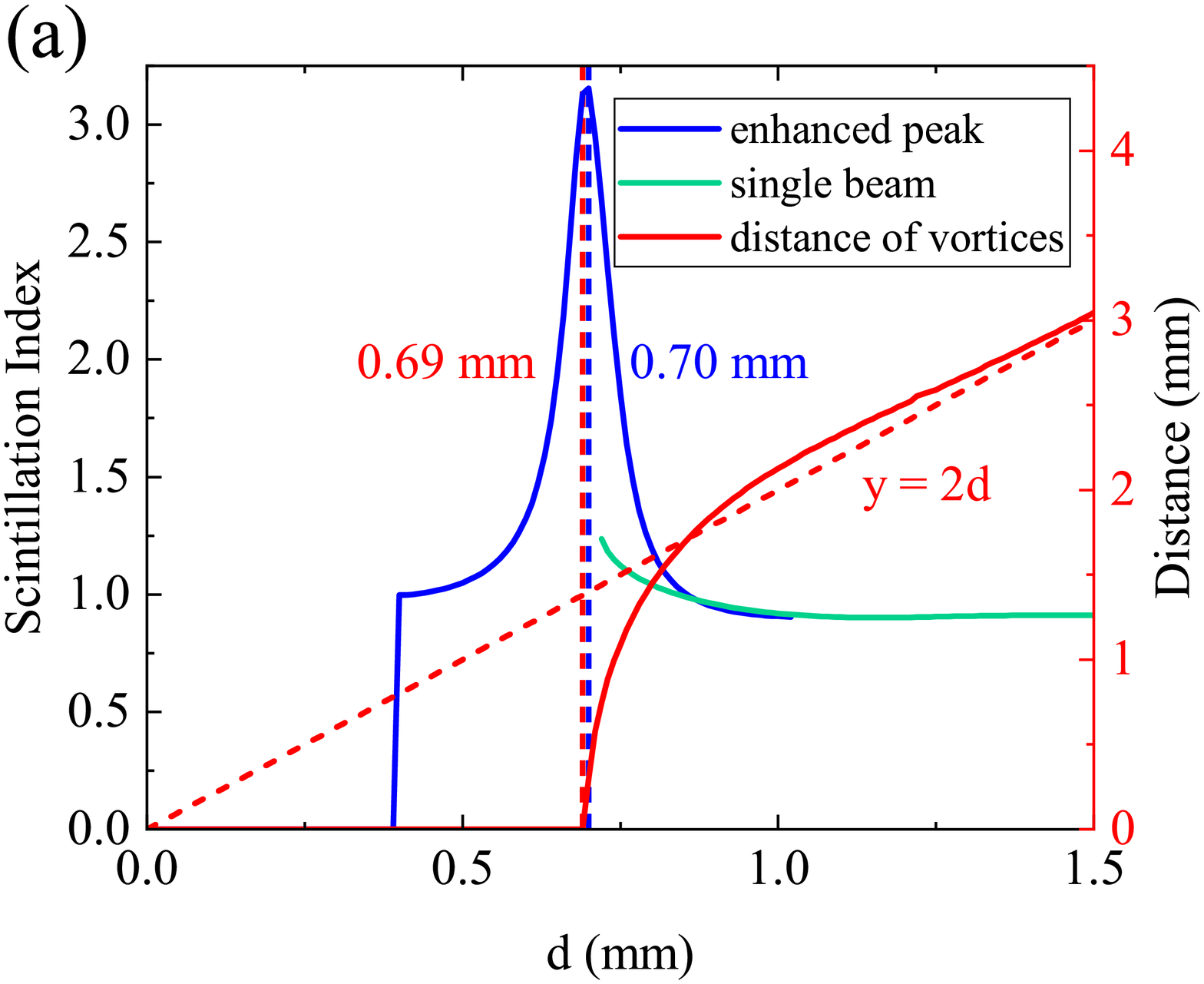}}\\
\subfloat{\includegraphics[width=\linewidth]{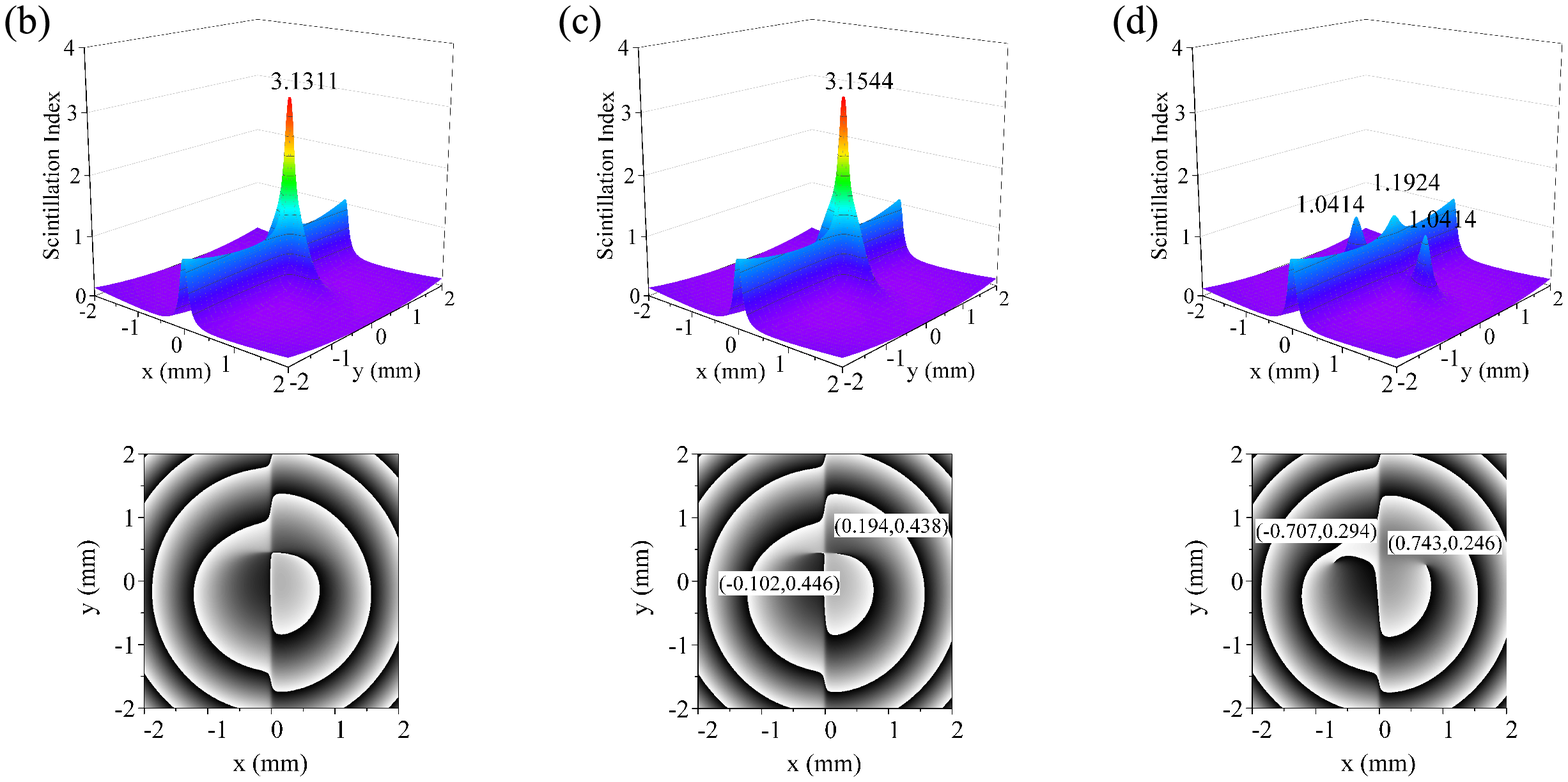}}
\caption{The evolution behaviors of the scintillation index of a beam with two types of phase discontinuities and the distance of two optical vortices the beam contains. (a) The local maximums of scintillation indexes and the distance of two screw dislocations for different values of $d$. The scintillation indexes and the phase distributions for (b) $d=0.69\rm{mm}$, (c) $d=0.70\rm{mm}$ and (d) $d=0.80\rm{mm}$.}\label{dchange}
\end{figure}

In this section, the transition state that screw dislocations and an infinitely extended edge dislocation coexist is investigated and the scintillation enhancement with detailed description is presented by graphs. The local maximums of scintillation indexes and the distance of two optical vortices are illustrated for different overlap ratios in Figure \ref{dchange}(a). To make it more intuitive, the scintillation index distributions and the phase distributions are shown for $d=0.69\rm{mm}$, $d=0.70\rm{mm}$ and $d=0.80\rm{mm}$ in Figs. 3(b)-3(d), respectively. Here we assume that the beam propagates for $5\rm{m}$ with $\sigma=1\rm{mm}$ and $\chi_{\rm{T}}$ on the transmission path is ${10^{ - 8}}{{\rm{K}}^2}{{\rm{s}}^{ - 1}}$. And the Rytov variance is ${2.6\times10^{-5}}$. With the increase of the separation distance of two Gaussian vortex beams, an enhanced peak of scintillation index appears on the scintillation index ridge. Then the peak value increases sharply to the maximum where $d=0.70\rm{mm}$. The further increase of the separation distance leads to the rapid drop of the scintillation index of the enhanced peak. Accompanied by the optical vortex keeping away from the scintillation index ridge, the enhanced peak gradually separates into two independent peaks of single Gaussian vortex beams as shown in Figure \ref{dchange}(d).

Focusing on the evolution behavior of the phase distribution, especially the screw dislocations, we find that the scintillation index has close connection with phase dislocations. Comparing Figure \ref{dchange}(b) and Figure \ref{dchange}(c), when the enhanced peak reaching the maximum value, two screw dislocations with opposite rotating directions abruptly appear at $\left( { - 0.102{\rm{mm}},0.446{\rm{mm}}} \right)$ and $\left( { 0.194{\rm{mm}},0.438{\rm{mm}}} \right)$, respectively. That is to say, the closer the screw dislocations to the infinitely extended edge dislocation are, the greater scintillation index will be. And the simultaneous appearance is accord with the character that topological charge is a conserved quantity under the influence of perturbations \cite{gbur2008vortex}. In addition, the distance of two screw dislocations increases rapidly and approaches to the length of $2d$ which is the separation distance of two Gaussian vortex beams. This trend is similar to the relation of intra-creation pair separation to propagation distance in Ref. \cite{2011Orbital}. The variation of the separation distance of two optical vortices reflects the attraction effect of the background field that caused by the infinitely extended edge dislocation indirectly. That is similar to the dynamical behavior of optical vortices in Ref. \cite{rozas1997propagation} and Ref.  \cite{roux1995dynamical}. 

\begin{figure}[htb]
\centering\includegraphics[width=0.4\linewidth]{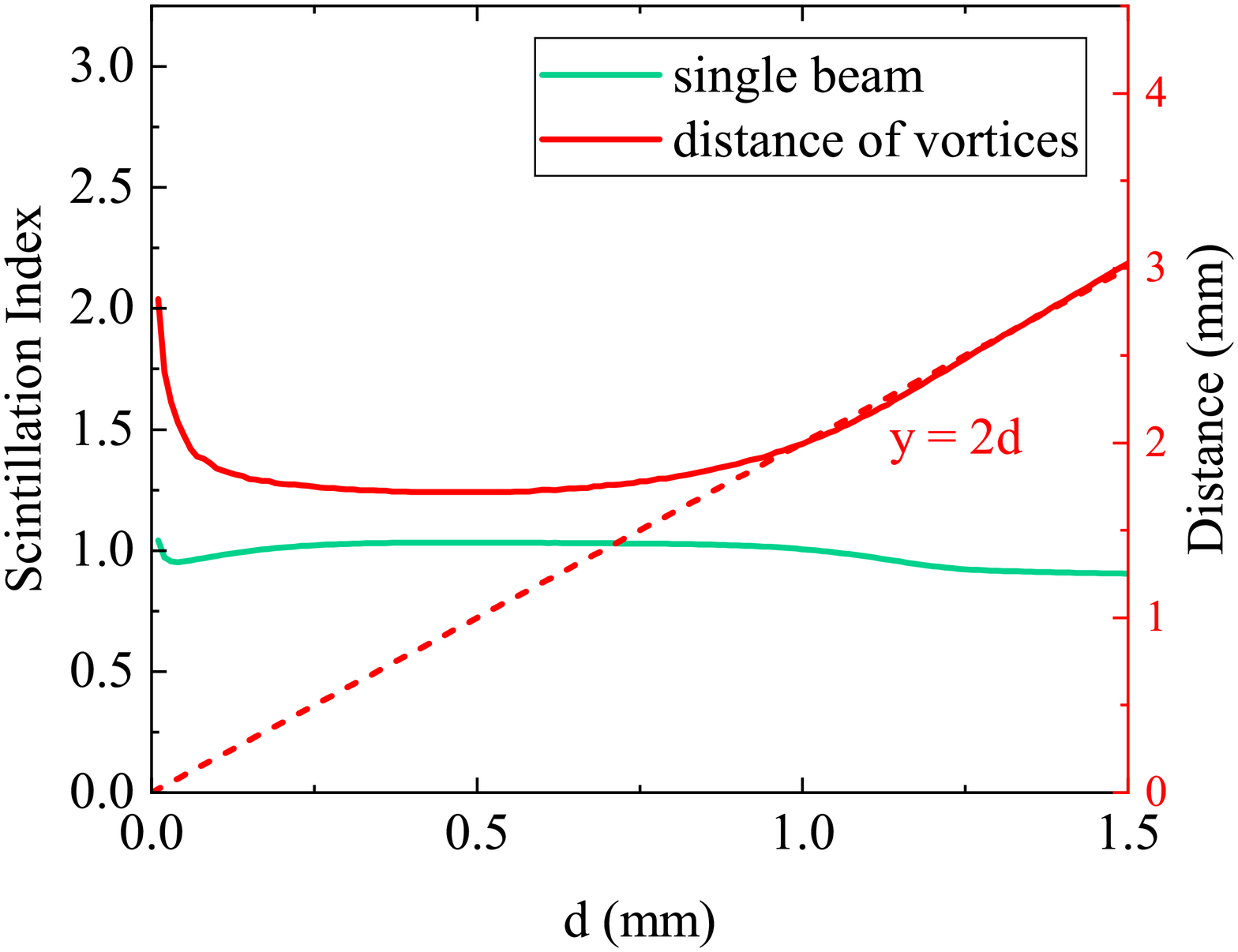}
\caption{The evolution behaviors of the local maximum of scintillation index and the distance of two optical vortices for only screw dislocations existing in a beam.}\label{dchangepi}
\end{figure}

To verify that the combination effect of two types of phase dislocations does induce the scintillation enhancement, we calculate the transmission results of the beam at receive plane with only screw dislocations existing. In a similar way, we plot the curves of the local maximums of scintillation indexes and the distance of two optical vortices when $\phi=\upi$ in Figure \ref{dchangepi}. 

The scintillation index of the single beam almost keeps invariant and has nearly the same value of that in Figure \ref{singlebeam}. When these two optical vortices are separated, the scintillation index peak mainly manifests as the independence of a single Gaussian vortex beam. But the variation of the distance of two optical vortices reflects a strong impact from the background field. The difference between the distance of vortices (red solid line in Figure \ref{dchangepi}) and the separation distance of two beams (red dashed line in Figure \ref{dchangepi}) indicates that the closer the vortices are, the stronger the repulsion effect will be. In fact, the change of the relative phase of two Gaussian vortex beams results in the disappearance of the infinitely extended edge dislocation. That makes the intensity gradient of the background field be the opposite state to the former case. And it alters the attraction effect to the repulsion effect on two optical vortices.

Another condition that only an infinitely extended edge dislocation existing has already been discussed in section 2. Therefore, through comparing the scintillation indexes in Figure \ref{doublebeam}, Figure \ref{dchange}(a) and Figure \ref{dchangepi}, the coexistence of screw dislocations and an infinitely extended edge dislocation results in the scintillation enhancement, and neither of these two types of dislocations can induce this phenomenon without the other.

There are two aspects mainly influencing the formation of scintillation enhancement. For the first aspect, the evolution behavior of optical vortices is the direct cause of scintillation enhancement. The spectral degree of coherent represents the expectation of the phase distribution which is shown in Figs. 3(b)-3(d). And it also charaterizes the probability of the annihilation of two optical vortices with opposite topological charges. For example, when ${d<0.70\rm{mm}}$ in Figure \ref{dchange}(a), there is no screw dislocation in the expectation of the phase distribution. But that does not mean there is no screw dislocation in any statistical sample. It just represents that the annihilation is in a dominant position. Therefore, a critical state must exist. And the creatiom amd the annihilation of optical vortices are evenly matched at this state. In addition, it is known that the zero amplitude points of a field are referred to as phase singularities \cite{2015Singular}. The creation or annihilation of optical vortices can cause obvious change in the intensity of the optical field in the neighborhood. And scintillation index changes when these two processes are nonnegligible under weak perturbations. Above all, the critical state may point to the state where scintillation index reaches the maximum, as shown in Figure \ref{dchange}(b) and Figure \ref{dchange}(c). As the state of the optical field moves away from the critical state, the enhanced scintillation fades away gradually. It can be seen clearly in Figure \ref{dchange}(a). Another aspect is that the background of optical field can influence the dynamical behavior of optical vortices \cite{ahluwalia2005evolution}. For example, the intensity valley induced by an infinitely extended edge dislocation in Figure \ref{dchange}(a). And the inverse example is shown in Figure \ref{dchangepi}. Besides, the variation of the background of an optical field due to oceanic turbulence has further influence on the dynamical behavior of optical vortices.
 
\begin{figure}[htb]
\centering\includegraphics[width=0.4\linewidth]{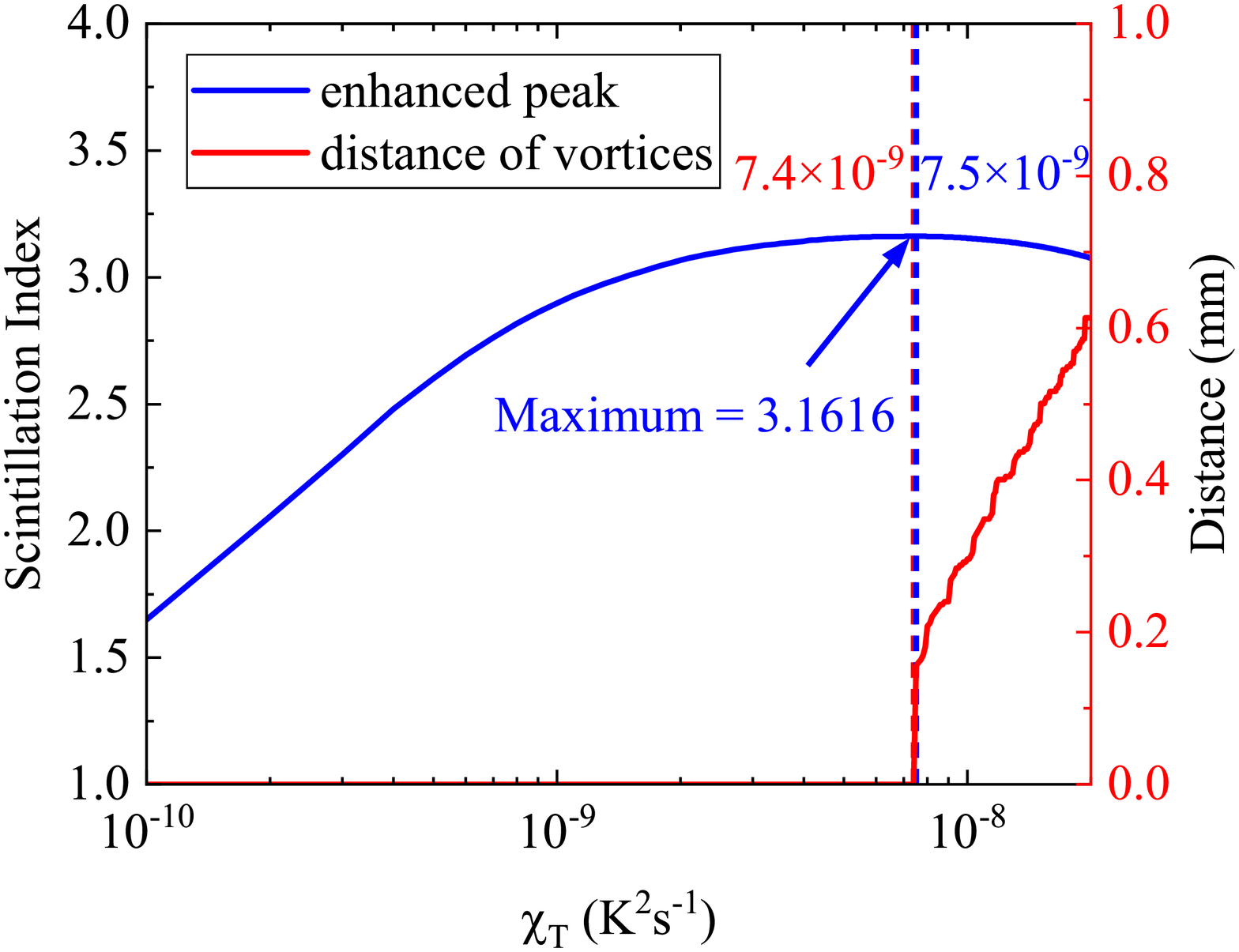}
\caption{The evolution behaviors of the local maximum of scintillation index and the distance of two optical vortices for only screw dislocations existing in a beam.}\label{kaiTchange}
\end{figure}

After verifying the formation condition of scintillation enhancement, the evolution behavior under the impact of oceanic turbulence for different intensity has been studied. The parameters of oceanic turbulence are set to be the same as the former except ${\chi}_{\rm{T}}$ varying from $10^{ - 10}{\rm{K}^2}{\rm{s}^{ - 1}}$ to ${2\times10^{ - 8}}{{\rm{K}}^2}{{\rm{s}}^{ - 1}}$. The $\sigma$ is still set to be $1\rm{mm}$ and the separation distance of two beams is selected to be $0.70\rm{mm}$ where the scintillation index of enhanced peak reaches the maximum (see blue dashed line in Figure \ref{dchange}(a)). Figure \ref{kaiTchange} plots the scintillation index of the enhanced peak and the distance of two vortices for different values of $\chi_{\rm{T}}$. It can be seen that the curve of scintillation index increases rapidly in the region of $\left[ {{{10}^{ - 10}},{{10}^{ - 9}}} \right]{{\rm{K}}^2}{{\rm{s}}^{ - 1}}$ and decelerates gradually before reaching the maximum value of $3.1616$ at ${\chi _{\rm{T}}} = 7.5 \times {10^{ - 9}}{{\rm{K}}^2}{{\rm{s}}^{ - 1}}$. Then, the curve begins to decrease with the appearance of optical vortices. We suppose the appearance of the independence of optical vortices is caused by the decrease of the steep degree of the intensity valley. That is to say, the evolution behavior of background optical field for different intensities of oceanic turbulence influences the behavior of optical vortices. And it is shown by the change of scintillation index. This saturation phenomenon is similar to the phenomenon in Figure \ref{dchange}(a) but not the scintillation saturation in strong fluctuations. 

Before reaching the saturation state, the variation of scintillation index is obviously larger than that for a single Gaussian vortex beam. And in this circumstance, the Rytov variance is in the region about the order of magnitude of $\left[ {{{10}^{ - 7}},{{10}^{ - 5}}} \right]$. In other words, the variation of scintillation index of the enhanced peak is about five orders of magnitude larger than that of a plane wave.

\section{The phase screen simulation of a beam with two types of phase discontinuities}

In section 3, the scintillation enhancement is presented in theory. To further verify this interesting phenomenon, in this section, we refer to the phase screen method in Ref. \cite{farwell2014multiple} to demonstrate the propagation process of the beam containing two types of phase discontinuities through weak oceanic turbulence. The parameters are set to be the same as that in Figure \ref{dchange}(a) for convenient comparation. For the setup of simulation, five phase screens with the size of $8 {\rm{mm}} \times 8 {\rm{mm}}$ are placed at intervals of   along the propagation path. Each screen has $512 \times 512$ points of analysis. And the split-step beam propagation method is used to simulate the propagation.

The evolution behavior of the scintillation index of the beam for different values of separation distance of two Gaussian vortex beams at receive plane is investigated. The distributions are shown in Figs. 6(a)-6(d). The scintillation index ridge of the region of the infinitely extended edge dislocation agrees with that plotting in Figure \ref{doublebeam}. And it is stable with the variation of the separation distance. The scintillation enhancement can be seen clearly in Figs. 6(b) and 6(c). The variation trend is almost the same as that in Figure \ref{dchange}(a) but with tiny difference. 

\begin{figure}[htb]
\centering\includegraphics[width=\linewidth]{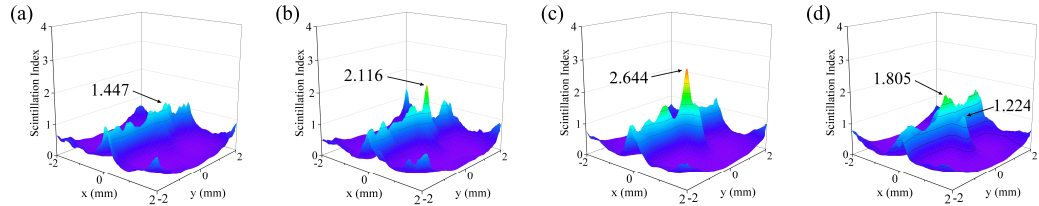}
\caption{The phase screen simulation results of the conditions in Figure \ref{dchange} for several values of the separation distance of two Gaussian vortex beams. (a) $d=0.60\rm{mm}$, (b) $d=0.65\rm{mm}$, (c) $d=0.70\rm{mm}$, (d) $d=0.75\rm{mm}$.}\label{phasescreen}
\end{figure}

Each case in Figure \ref{phasescreen} is the ensemble average of $500$ realizations because the further increase of realizations has less impact on the error reduction. According to the scintillation index of enhanced peak in Figure \ref{dchange}(a), the error is within $16.1\% $. That is mainly due to the sensitivity of the scintillation index peak to the oceanic turbulence structure whose scale is around or smaller than the transverse scale of the beam. This perturbation leads to the unstability of the scintillation index peak and it is obvious in Figure \ref{phasescreen}(c) that the peak value has great difference from that in Figure \ref{dchange}(c). However, the fluctuation of scintillation index is not strong enough to hinder the observation of scintillation enhancement. Because this phenomenon mainly caused by the oceanic turbulence structure with the scale larger than the transverse scale of the beam. 

In a word, the prediction of scintillation enhancement in analytical derivation can be validated in this section.

\section{Conclusion}

In summary, we have investigated the propagation behavior of a beam consisted by two coherent Gaussian vortex beams with ${\pm{1}}$ topological charges through weak oceanic turbulence under some approximations. By means of varying the parameters of the optical field, the scintillation indexes for different combination conditions of screw dislocations and an infinitely extended edge dislocation have been calculated. Among these cases, the scintillation enhancement induced by an infinitely extended edge dislocation and screw dislocations has been verified in theory by comparing the condition of coexistence with the condition of the existence of any one of two types of phase dislocations. Furthermore, the analytical derivation has been validated by the phase screen simulation. 

The scintillation enhancement provides a new point to observe the dynamical behavior of phase dislocations, especially the beam propagating through random media. For further research on strong turbulence, it has potential application value in characterizing strong turbulence from branch points detection. Furthermore, its high sensitivity for weak fluctuation opens up new possibilities for the measurement of weak oceanic turbulence. 

\bibliographystyle{tfnlm}
\bibliography{sample2}

\begin{thebibliography}{10}
\providecommand{\url}[1]{\normalfont{#1}}
\providecommand{\urlprefix}{Available from: }

\bibitem{nye1974dislocations}
Nye~JF, Berry~MV. Dislocations in wave trains. Proceedings of the Royal Society
  of London A Mathematical and Physical Sciences.
  1974;\hspace{0pt}336(1605):165--190.

\bibitem{fried1992branch}
Fried~DL, Vaughn~JL. Branch cuts in the phase function. Applied Optics.
  1992;\hspace{0pt}31(15):2865--2882.

\bibitem{basistiy1995optical}
Basistiy~I, Soskin~M, Vasnetsov~M. Optical wavefront dislocations and their
  properties. Optics Communications. 1995;\hspace{0pt}119(5-6):604--612.

\bibitem{lugiato1990instabilities}
Lugiato~L, Oppo~G, Tredicce~J, et~al. Instabilities and spatial complexity in a
  laser. JOSA B. 1990;\hspace{0pt}7(6):1019--1033.

\bibitem{coullet1989optical}
Coullet~P, Gil~L, Rocca~F. Optical vortices. Optics Communications.
  1989;\hspace{0pt}73(5):403--408.

\bibitem{ahluwalia2005evolution}
Ahluwalia~B, Yuan~XC, Tao~S. Evolution of composite off-axis vortexes embedded
  in the propagation-invariant beams. Optics Communications.
  2005;\hspace{0pt}247(1-3):1--9.

\bibitem{cheng2016propagation}
Cheng~M, Guo~L, Li~J, et~al. Propagation of an optical vortex carried by a
  partially coherent laguerre--gaussian beam in turbulent ocean. Applied
  Optics. 2016;\hspace{0pt}55(17):4642--4648.

\bibitem{ren2016orbital}
Ren~Y, Li~L, Wang~Z, et~al. Orbital angular momentum-based space division
  multiplexing for high-capacity underwater optical communications. Scientific
  Reports. 2016;\hspace{0pt}6:33306.

\bibitem{2016Vortex}
Soifer~VA, Korotkova~O, Khonina~SN, et~al. Vortex beams in turbulent media:
  Review. University of Miami. 2016;\hspace{0pt}40(5):605--624.

\bibitem{li2017influence}
Li~Y, Yu~L, Zhang~Y. Influence of anisotropic turbulence on the orbital angular
  momentum modes of hermite-gaussian vortex beam in the ocean. Optics Express.
  2017;\hspace{0pt}25(11):12203--12215.

\bibitem{liu2013experimental}
Liu~X, Shen~Y, Liu~L, et~al. Experimental demonstration of vortex phase-induced
  reduction in scintillation of a partially coherent beam. Optics Letters.
  2013;\hspace{0pt}38(24):5323--5326.

\bibitem{voitsekhovich1998density}
Voitsekhovich~VV, Kouznetsov~D, Morozov~DK. Density of turbulence-induced phase
  dislocations. Applied Optics. 1998;\hspace{0pt}37(21):4525--4535.

\bibitem{rao2008statistics}
Rao~R. Statistics of the fractal structure and phase singularity of a plane
  light wave propagation in atmospheric turbulence. Applied Optics.
  2008;\hspace{0pt}47(2):269--276.

\bibitem{oesch2009aggregate}
Oesch~DW, Sanchez~DJ, Tewksbury-Christle~CM, et~al. The aggregate behavior of
  branch points: branch point density as a characteristic of an atmospheric
  turbulence simulator. In: Advanced Wavefront Control: Methods, Devices, and
  Applications VII; Vol. 7466; International Society for Optics and Photonics;
  2009. p. 746606.

\bibitem{2009The}
Sanchez~DJ, Oesch~DW. The aggregate behavior of branch points: the creation and
  evolution of branch points. In: Advanced Wavefront Control: Methods, Devices,
  and Applications VII; 2009.

\bibitem{he1995direct}
He~H, Friese~M, Heckenberg~N, et~al. Direct observation of transfer of angular
  momentum to absorptive particles from a laser beam with a phase singularity.
  Physical Review Letters. 1995;\hspace{0pt}75(5):826.

\bibitem{vadnjal2013measurement}
Vadnjal~AL, Etchepareborda~P, Federico~A, et~al. Measurement of in-plane
  displacements using the phase singularities generated by directional wavelet
  transforms of speckle pattern images. Applied Optics.
  2013;\hspace{0pt}52(9):1805--1813.

\bibitem{aksenov2012increase}
Aksenov~VP, Pogutsa~CE. Increase in laser beam resistance to random
  inhomogeneities of atmospheric permittivity with an optical vortex included
  in the beam structure. Applied Optics. 2012;\hspace{0pt}51(30):7262--7267.

\bibitem{2011Orbital}
Sanchez~DJ, Oesch~DW. Orbital angular momentum in optical waves propagating
  through distributed turbulence. Optics Express.
  2011;\hspace{0pt}19(24):24596--608.

\bibitem{freund1999critical}
Freund~I. Critical point explosions in two-dimensional wave fields. Optics
  Communications. 1999;\hspace{0pt}159(1-3):99--117.

\bibitem{andrews2005laser}
Andrews~LC, Phillips~RL. Laser beam propagation through random media. SPIE;
  2005.

\bibitem{huang2014evolution}
Huang~Y, Zhang~B, Gao~Z, et~al. Evolution behavior of gaussian schell-model
  vortex beams propagating through oceanic turbulence. Optics Express.
  2014;\hspace{0pt}22(15):17723--17734.

\bibitem{Shirai2003Mode}
Shirai, Tomohiro, Dogariu, et~al. Mode analysis of spreading of partially
  coherent beams propagating through atmospheric turbulence. JOSA A.
  2003;\hspace{0pt}20(6):1094--1102.

\bibitem{yao2019wide}
Yao~JR, Zhang~HJ, Wang~RN, et~al. Wide-range prandtl/schmidt number power
  spectrum of optical turbulence and its application to oceanic light
  propagation. Optics Express. 2019;\hspace{0pt}27(20):27807--27819.

\bibitem{yao2020spatial}
Yao~JR, Elamassie~M, Korotkova~O. Spatial power spectrum of natural water
  turbulence with any average temperature, salinity concentration, and light
  wavelength. JOSA A. 2020;\hspace{0pt}37(10):1614--1621.

\bibitem{gradshteyn2014table}
Gradshteyn~IS, Ryzhik~IM. Table of integrals, series, and products. Academic
  press; 2014.

\bibitem{mandel1995optical}
Mandel~L, Wolf~E. Optical coherence and quantum optics. Cambridge university
  press; 1995.

\bibitem{gbur2003coherence}
Gbur~G, Visser~TD. Coherence vortices in partially coherent beams. Optics
  Communications. 2003;\hspace{0pt}222(1-6):117--125.

\bibitem{aksenov2015scintillations}
Aksenov~VP, Kolosov~VV. Scintillations of optical vortex in randomly
  inhomogeneous medium. Photonics Research. 2015;\hspace{0pt}3(2):44--47.

\bibitem{gbur2008vortex}
Gbur~G, Tyson~RK. Vortex beam propagation through atmospheric turbulence and
  topological charge conservation. JOSA A. 2008;\hspace{0pt}25(1):225--230.

\bibitem{rozas1997propagation}
Rozas~D, Law~C, Swartzlander~G. Propagation dynamics of optical vortices. JOSA
  B. 1997;\hspace{0pt}14(11):3054--3065.

\bibitem{roux1995dynamical}
Roux~FS. Dynamical behavior of optical vortices. JOSA B.
  1995;\hspace{0pt}12(7):1215--1221.

\bibitem{2015Singular}
Gbur~G. Singular optics. American Cancer Society; 2015.

\bibitem{farwell2014multiple}
Farwell~NH, Korotkova~O. Multiple phase-screen simulation of oceanic beam
  propagation. In: Laser Communication and Propagation through the Atmosphere
  and Oceans III; Vol. 9224; International Society for Optics and Photonics;
  2014. p. 922416.

\end{thebibliography}

\section*{Disclosure statement}

No potential conflict of interest was reported by the author(s).


\end{document}